\documentclass[aps,
prd,
preprint,
eqsecnum,
amsmath,
amssymb
]{revtex4}

\usepackage{hyperref}
\usepackage{color}
\usepackage{mathrsfs}
\usepackage{xcolor}

\begin{document}

\title{Hamiltonian BRST formalism for gauge fields on black hole spacetimes}

\author{Karan Fernandes} \email[email: ]{kfernandes@cmi.ac.in}
\affiliation{Chennai Mathematical Institute, SIPCOT IT Park, Siruseri, Chennai 603103, India}

\author{Amitabha Lahiri} \email[email: ]{amitabha@boson.bose.res.in}
\affiliation{S N Bose National Centre for Basic Sciences, Block JD, Sector III, Salt Lake, Kolkata 700106, India}

\begin{abstract}
We investigate the Becchi-Rouet-Stora-Tyutin (BRST) formalism for gauge theories on spherically symmetric black hole spacetimes, 
with or without a cosmological constant ($\Lambda\geq0$). This is illustrated through the example of 
scalar electrodynamics. We first demonstrate that the horizons contribute additional surface terms 
to the Gauss law constraint of the theory when gauge transformations are not required to vanish on the 
horizons. We then consider the BRST invariant path integral including these surface terms 
following the Hamiltonian BRST formalism. We fix a radiation-like gauge which involves null components 
of the electromagnetic field at the horizons. We find that the presence of the surface terms in the constraint
forces the ghost and gauge fixing actions to include additional terms at the horizons. 
The null combination of the gauge fields at the horizons is shown to modify the ghost number charge of the 
theory through additional terms at the horizons. We also demonstrate how one can construct a gauge-fixing 
fermion which generates its own nilpotent symmetry transformations, called co-BRST transformations, that 
leave the theory invariant. The BRST and co-BRST transformations are further used to identify dressed 
(gauge invariant) fields of theory, whose dressings are affected by the presence of Killing horizons. 
We conclude with a discussion of potential applications of our results in soft limits and thermal field 
theories on black hole backgrounds.

\end{abstract}


\maketitle

\section{Introduction}

Gauge field theories play a central role in our understanding 
of interactions between elementary particles. The reasons behind 
their usefulness as quantum field theories valid to very short 
distances are renormalizability and unitarity. The Becchi-Rouet-Stora-Tyutin 
(BRST) symmetry~\cite{Becchi:1975nq, Tyutin:1975qk} provides a very 
convenient tool for verifying these properties of 
a gauge theory. While any theory with a hermitian Hamiltonian operator is 
necessarily unitary, a gauge field theory has redundant degrees of freedom 
which have to be eliminated by gauge-fixing. Apart from a few exceptions, 
gauge-fixing terms introduce states of negative norm into the theory,
which in turn must be eliminated by the introduction of ghost fields. 
A test of unitarity of a gauge theory is to verify if the action, including
the gauge-fixing and ghost terms, is invariant under BRST symmetry. 
The corresponding conserved charge $Q_{BRST}$ is nilpotent and defines
a cohomology on the Fock space of the theory, leading to a consistent 
separation of physical and unphysical states, and the theory is unitary
on the physical subspace. Since the redundancy of gauge theories is
manifested in the constraints, one way of constructing the BRST charge is to 
start with the constrained Hamiltonian of the theory and introduce
ghost fields and their conjugate momenta as Lagrange multipliers for 
the constraints in the BRST charge~\cite{Henneaux:1985kr,Henneaux:1992ig}.
In this paper we follow this route for gauge theories on black hole 
spacetimes, paying close attention to the effect of horizons.

It is known that the presence of spatial boundaries 
on the manifold can significantly modify the dynamics and quantization of 
gauge theories. However, the consequences on gauge fields arising from spatial 
and null boundaries can significantly differ due to the properties of gauge 
fields and the underlying symmetries of these surfaces. For instance, it has been 
recently understood that the infinite-dimensional symmetry groups of gauge and 
gravitational fields at null infinity $\mathscr{I}$ on asymptotically flat 
spacetimes imply the existence of an infinite number of soft charges on the 
sphere at null infinity~\cite{He:2014cra,He:2014laa,Kapec:2015ena}. 
Of central importance to these results are the allowed non-vanishing gauge 
parameters which depend only on the angular variables. The presence of soft 
charges on the sphere at null infinity, along with the requirement of charge 
conservation, has been used to argue that soft electromagnetic and gravitational 
hairs should also exist on the horizons of black holes on asymptotically 
flat backgrounds~\cite{Hawking:2016msc,Hawking:2016sgy}. These results have further motivated 
the investigation of conserved charges and currents on null surfaces in 
general~\cite{Hopfmuller:2018fni,Chandrasekaran:2018aop,Blau:2015nee}. Thus gauge and 
gravitational fields, or more generally constrained field theories, might 
provide further insight into our understanding of black holes. 

Constrained field theories can be understood within the well established 
Dirac-Bergmann and BRST formalisms. Ordinarily in the case 
of curved backgrounds with spatial boundaries, surface terms are introduced 
in the Hamiltonian to provide constraints without any surface 
contributions~\cite{Regge:1974otg,Regge:1974zd,Benguria:1976in}. This is 
consistent with the requirement that gauge parameters vanish at spatial infinity. More generally, 
the regularity of fields, 
including gauge fields, can be invoked to fix the parameters of gauge 
transformations at spatial boundaries~\cite{Balachandran:1993tm,Balachandran:1992qg,Balachandran:1995dv}. 
As a consequence, the constraints of the theory involve no corrections 
from the boundary and any surface terms which arise from the Dirac-Bergmann 
formalism are identified with additional boundary conditions which must be 
imposed on the fields~\cite{Isenberg:1981fa,SheikhJabbari:1999xd,Zabzine:2000ds}.

In the case of BRST invariant actions on manifolds with spatial boundaries, boundary 
conditions on the gauge fields imply a general class of boundary conditions 
on the ghosts in order to ensure the BRST invariance of the 
theory~\cite{Moss:1989wu,Moss:1990yq,Moss:1996ip,Moss:2013vh}. There have 
also been recent considerations of spatial bounding surfaces within the 
manifold, as in the case of entagling surfaces used to investigate the 
entanglement entropy. While the consideration of edge modes and boundary 
conditions on the gauge fields have important implications on entanglement 
entropy calculations~\cite{Donnelly:2013tia,Donnelly:2014fua,Donnelly:2015hxa,
Nishioka:2018khk,Blommaert:2018rsf,Blommaert:2018oue,Barnich:2018zdg}, in 
all known cases with spatial boundaries the constraints of gauge theories 
are not modified by the presence of spatial boundaries. 
%
%


The situation is different for Killing horizons, such as black hole 
event horizons or cosmological horizons. These are not physical boundaries even though
the (timelike) Killing vector field becomes null on these 3-surfaces. Specifically,
this implies that fields need not be set to zero on horizons, and therefore 
the parameters of gauge transformations need not vanish on 
these `null boundaries' as 
they do on spatial boundaries. In the case of null and spatial infinities, which 
represent the boundaries of a compactified manifold, we can however identify 
fall-off conditions on the fields defined on the background which can help restrict 
the behaviour of gauge fields there. This is not the case for Killing horizons, 
which can represent globally defined null surfaces on black hole backgrounds. 
This includes the case 
of black hole de Sitter backgrounds, where the black hole event horizon and 
cosmological horizon are finitely located within the global manifold.  As such, 
we cannot a priori impose any conditions on gauge fields at Killing horizons. 
We can only insist that gauge invariant scalars, such as those appearing in 
the stress tensor, are finite at the horizons. 

The finiteness of gauge-invariant scalars and the arbitrariness of gauge parameters 
at the horizons served as the basis of our recent work on the constrained dynamics 
of field theories on curved backgrounds with Killing horizons~\cite{Fernandes:2016imn,Fernandes:2018pii}. 
Using the Dirac-Bergmann formalism, it was shown there that the Gauss law constraint of gauge theories gets
additional surface contributions from the horizons of the background. This modifies
the charges and also allows for interesting gauge fixing choices involving surface terms at the horizons.
In this paper, we continue our investigation of constrained field theories on curved backgrounds 
with horizons, now using the Hamiltonian BRST formalism. 
In particular, using the example of scalar electrodynamics, 
we will investigate the effect of horizons on interacting gauge theories. We will 
find that like the Gauss law constraint, the BRST charge will also pick up a horizon contribution. 
While we do not consider all possible implications of our results in this paper, we will 
consruct a co-BRST operator and use the invariance under BRST and co-BRST to identify dressed
fields as in~\cite{Lavelle:1993xf}. In our work, the spacetime
is treated as a fixed background; we have not considered gravitational constraints.

We first demonstrate, using the Dirac-Bergmann formalism, that the Gauss law constraint 
of the theory receives additional surface contributions from the black hole horizon and 
also from the cosmological horizon, if it is present. We then extend the phase space and 
consider the Hamiltonian BRST formalism for the theory on spherically symmetric backgrounds 
with horizons. The BRST charge inherits the horizon terms which are present in the Gauss 
law constraint of the theory. The horizon terms in the Gauss law constraint ensure that 
BRST transformations have their usual expressions on curved backgrounds without boundaries. 
However, the BRST transformations which we derive also act on fields at the horizons. Thus 
in general, we require a gauge with surface terms to fix the theory at the horizons of the 
spacetime. For fixing the gauge, we use a modified radiation gauge which include null 
components of the electromagnetic field at the horizons. This choice leads to surface 
integrals at the horizons in the BRST invariant action and a ghost number charge which involves 
additional terms from the horizons of the background. 

The gauge fixing fermion in the Hamiltonian BRST formalism can be chosen such that it is 
a generator of a new set of nilpotent symmetry transformations, different from the BRST 
transformations, which leave the theory invariant. These are known as co-BRST transformations, 
which have been considered for scalar electrodynamics in flat 
spacetime~\cite{Lavelle:1993xf,Marnelius:1997by}. Invariance under both BRST and co-BRST 
transformations help identify dressed scalar fields as physical fields in flat spacetime. 
In the context of the present work, we find an additional contribution to the dressing 
arising from the horizons. We conclude our paper with a discussion 
on the potential implications of our results on soft limits at the horizons of black 
holes and thermal field theories on black hole spacetimes. 

The organization of our paper is as follows. In Sec.~\ref{geom} we set up our notations and 
conventions for gauge theories on spherically symmetric backgrounds with horizons. 
In Sec.~\ref{SQED} we consider scalar electrodynamics on spacetimes with horizons 
and show that the Gauss law constraint has to include an additional surface term. 
We also discuss consistent gauge fixing choices which involve surface terms at the horizons,
to be used in the Hamiltonian BRST formalism. In Sec.~\ref{BRS}, we describe the 
the derivation of the BRST path integral from the Hamiltonian and derive 
the BRST invariant action in a gauge which involve null components of the gauge 
field at the horizons. We show that the ghost number charge of the theory 
in this gauge involves additional terms at the horizons of the background. 

In Sec.~\ref{sec.cbrst}, we define a gauge fixing fermion which generates nilpotent 
co-BRST transformations that leave the theory invariant. The BRST and co-BRST 
transformations are used to identify the dressing of scalar fields of the theory 
on non-asymptotically flat backgrounds. In Sec.~\ref{Con}, we describe how our 
results could be used to further investigate infrared limits and thermal gauge 
theories on black hole backgrounds.

\section{Geometric Framework} \label{geom}
We begin by considering some essential preliminaries needed for the remaining sections. 
The BRST construction will be carried out on a static, spherically symmetric and 
torsion-free manifold $\mathcal{M}$ endowed with at least one horizon. 
In other words, we only assume that the spacetime possesses a timelike Killing vector field 
$\xi^a\,$ normalized as $\xi^a\xi_a = -\lambda^2\,,$ which satisfies 
\begin{align}
\xi_{[a}\nabla_b \xi_{c ]} & = 0 \,. \label{Frobenius} 
\end{align}
It follows that there exists a spacelike hypersurface $\Sigma$ which is 
everywhere orthogonal to $\xi^a\,.$ The horizon is defined by $\xi^a$ becoming null, 
$\lambda=0\,.$  For an asymptotically flat or anti-de Sitter space, $\Sigma$ is the 
region `outside the horizon', while for backgrounds with a positive cosmological constant, 
such as static de Sitter black hole spacetimes, $\Sigma$ is the region `between the 
horizons'. 

The induced metric $h_{ab}$ and projection operator $h^a_b$ on $\Sigma$ are given by
\begin{equation}
h_{ab} = g_{ab} + {\lambda}^{-2} \xi_{a} \xi_{b}\,, \quad 
h^a_b = \delta^a_b + \lambda^{-2}\xi^a \xi_b \,.
\label{gen.met}
\end{equation}
leading to the following expression for the determinant of spacetime metric
\begin{equation}
\sqrt{-g} = \lambda \, \sqrt{h} \, . \label{gen.det}
\end{equation}
We will denote the Killing horizons of the spacetime as $\mathcal{H}$.
The intersection of $\mathcal{H}$ with $\Sigma$ is topologically a 2-sphere
	(or the union of two 2-spheres, if there is a cosmological horizon) with an induced 
metric $\sigma_{ab}$ which can be written as 
\begin{equation}
\sigma_{ab} = h_{ab} - n_a n_b \,,
\label{hor.proj}
\end{equation}
 where $n^a$ is the outward (inward) pointing unit spatial normal to the inner (outer) horizon, 
which points into $\Sigma$ and satisfies $n_a n^a = 1$\,. We will also refer to these 2-spheres as 
the `horizon' and write them as $\partial\Sigma$.  Note that $\partial\Sigma$ is not a physical boundary 
space in any sense, fields or their functions do not need to vanish or diverge there in general. 
In the context of gauge theories, we only require gauge invariant scalars constructed out of the 
fields to be finite on $\partial \Sigma$\,.

If we need to refer to $\mathcal{H}\,,$ which is of course null, we will call it the spacetime horizon.
 In this paper we will also make use of the null normals of $\mathcal{H}\,,$ 
defined in spacetime. On the spherically symmetric backgrounds we are considering, we can define the two
null normals $l_a$ and $k_a$ in terms of the normalized timelike Killing vector field $\lambda^{-1}\xi_a$ 
and the unit spatial normal $n_a$ to the spatial sections of the null hypersurface,
\begin{equation}
l_a = \frac{1}{\sqrt{2}}\left(\lambda^{-1} \xi_a + n_a \right) \,, \qquad k_a = \frac{1}{\sqrt{2}}\left(\lambda^{-1} \xi_a - n_a \right) \,.
\label{nn.ss}
\end{equation}
For this choice of the null normals, we find that 
$l_a$ and $k_a$  satisfy
\begin{equation}
l_a l^a = 0 = k_ak^a \,, \qquad  l_a k^a = -1 \,,
\label{nn}
\end{equation} 
and we can write the metric on the null hypersurface as 
\begin{equation}
\tilde{\sigma}_{ab} = g_{ab} + l_{a} k_{b} + l_{a} k_{b} \,.
\label{hor.proj2}
\end{equation}
These expressions hold for all null hypersurfaces of the background, including $\cal{H}$.

The BRST formalism requires fields which belong to the Grassmann algebra, of both even 
and odd Grassmann parity. Denoting Grassmann parity by $\epsilon$, we say that the field 
$\Phi_A$ is `even' when $\epsilon_{\Phi_A} = 0$ (mod $2$) and that it is `odd' when 
$\epsilon_{\Phi_A} = 1$ (mod $2$). Lagrangians and Hamiltonians will always be an even 
functional of the fields. Because Grassmann parity is additive for composite fields, given any 
two functionals of the fields $F(\Phi_A)$ and $G(\Phi_A)$, we have
\begin{equation} 
F G = (-1)^{\epsilon_F \epsilon_G} G F \, .
\label{g.gras}
\end{equation}
Due to the presence of odd Grassmanian fields, the derivatives of functionals have to be 
handled carefully. The derivative of a functional $F(\Phi_A)$ of a field $\Phi_A$ can be written in two possible ways
\begin{equation}
\text{either} \quad  \frac{\delta_L F}{\delta \Phi_A} \, , \quad \text{or} \,\quad  \frac{\delta_R F}{\delta \Phi_A}  \, ,
\label{gen.Grd}
\end{equation}
where $\displaystyle{\frac{\delta_L }{\delta \Phi_A}}$ and $\displaystyle{\frac{\delta_R }
{\delta \Phi_A}}$ denote the left and right functional derivatives with respect to $\Phi_A$,
respectively. For the left functional derivative $\displaystyle{\frac{\delta_L }{\delta \Phi_A}}\,,$ 
we vary $F$ with respect to $\Phi_A$, with $\delta \Phi_A$ 
moved to the extreme left using Eq.~(\ref{g.gras}) and then deleted. Likewise the right 
functional derivative $\displaystyle{\frac{\delta_R }{\delta \Phi_A}}$ means that $F$ is 
varied with respect to $\Phi_A$, with $\delta \Phi_A$ moved to the extreme right and 
then deleted. 
These derivatives are identical when the field $\Phi_A$ is even. In the following, functional 
derivatives will always be taken to mean `left' unless specified otherwise.

The action functional for $N$ fields $\Phi_A\,,\, A = 1, \cdots, N\,,$ is given by the time 
integral of the Lagrangian $L$
\begin{equation}
S[\Phi_A] = \int dt\, L  
= \int dt \int \limits_{\Sigma} dV_x ~ {\cal L}(\Phi_A(x) , \nabla_a \Phi_A(x)) \, ,
\label{gen.act}
\end{equation}
where $dV_x$ is the volume element on $\Sigma\,,$ 
and ${\cal L}(\Phi_A(x) , \nabla_a \Phi_A(x))$ is the Lagrangian density. The Lagrangian density 
can be written in terms of the `spatial' and `temporal' derivatives of the fields, 
\begin{equation}
{\cal L} \equiv {\cal L}(\Phi_A(x),  \mathcal{D}_a \Phi_A(x),  \dot{\Phi}_A(x) )\,,
\label{gen.den}
\end{equation}
where $\mathcal{D}_a \Phi_A = h_a^b \nabla_b \Phi_A$ are the $\Sigma$-projected derivatives 
of the fields $\Phi_A\,,$ and $\dot{\Phi}_A$ are their time derivatives, defined as their Lie derivatives
with respect to $\xi\,,$
\begin{equation}
\dot{\Phi}_A := \pounds_{\xi} \Phi_A \,.
\label{gen.dot}
\end{equation}
The momenta $\Pi^A$ canonically conjugate to the fields $\Phi_A\,$ are defined as
\begin{equation}
\Pi^A  = \frac{\delta L}{\delta \dot{\Phi}_A} \, ,
\label{gen.mom}
\end{equation}
where the functional derivative in this definition is taken on the hypersurface $\Sigma\,,$ i.e. 
it is an `equal-time' functional derivative, defined as
\begin{equation}
\frac{\delta\Phi_A(\vec{x}, t)}{\delta\Phi_B(\vec{y}, t)} = \delta^B_A\, \delta(x, y) = 
\frac{\delta\dot\Phi_A(\vec{x}, t)}{\delta\dot\Phi_B(\vec{y}, t)}\, .
\label{gen.var}
\end{equation}
The $\delta(x,y)$ in Eq.~(\ref{gen.var}) is the three-dimensional covariant delta function defined 
on $\Sigma\,,$
\begin{equation}
\int\limits_\Sigma dV_y \delta(x,y) f(\vec{y}, t) = f(\vec{x}, t)\,.
\label{gen.del}
\end{equation}
Given a Lagrangian $L$ we can construct the canonical Hamiltonian through the Legendre transform
\begin{equation}
H_C = \int \limits_{\Sigma} dV_x ~(\Pi^A \dot{\Phi}_A) -  L \,.
\label{gen.Ham}
\end{equation}
The generalized Poisson bracket for two functionals $F\left(\Phi_A, \Pi^A\right)$ and 
$G\left(\Phi_A, \Pi^A\right)$ is defined as
\begin{align}
\left[F ,G\right]_P  = \int dV_z \left(\frac{\delta_R F}{\delta \Phi_A(z)} \frac{\delta_L G}{\delta \Pi^A(z)} 
- \frac{\delta_R F}{\delta \Pi^A(z)} \frac{\delta_L G}{\delta \Phi_A(z)} \right)\,. \label{gen.PB}
\end{align}
In accounting for Grassmann fields, the generalized Poisson bracket reduces to a 
commutator when any one of the fields is even and an anticommutator when both fields 
are odd. We will henceforth refer to this bracket simply as the Poisson bracket.
With the choice of $F(x) = \Pi^B(\vec{x}, t)$ and $G(y) = \Phi_A(\vec{y}, t)$, we 
recover the canonical relation between the fields and their momenta
\begin{equation}
\left[\Pi^B(\vec{x}, t) , \Phi_A(\vec{y}, t)\right]_P = - \delta_{A}^{B} \delta(x,y)\,. 
\label{gen.can}
\end{equation}
The time evolution of any functional of the fields can also be determined from its Poisson bracket with the Hamiltonian.
\begin{equation}
\dot{F}(x) = \left[F(x), H_C \right]_P\,.
 \end{equation}
Using the above definitions, we can now consider Hamiltonian formalisms for constrained 
field theories on spherically symmetric backgrounds with horizons. There exist several 
excellent textbooks and reviews which cover these topics (see for example~\cite{Dirac-lect-1964, 
Henneaux:1992ig, Sundermeyer:1982gv, Hanson:1976cn, Mukunda:1974dr} for the Dirac-Bergmann 
formalism and~\cite{Henneaux:1985kr, Henneaux:1992ig} for the Hamiltonian BRST approach), 
and we will not review them here.
%

\section{Scalar Electrodynamics }\label{SQED}
Before proceeding to the Hamiltonian BRST treatment in the next section, we will first 
need to identify all the constraints of the theory using the Dirac-Bergmann formalism in 
order to define the BRST charge operator. The procedure will follow the standard treatment 
in~\cite{Dirac-lect-1964, Henneaux:1992ig, Sundermeyer:1982gv, Hanson:1976cn, Mukunda:1974dr}. 
In the case of scalar electrodynamics, the results are simply curved spacetime 
generalizations of those known in flat spacetime, with the exception of the form 
of the Gauss law constraint. As promised earlier, we will find that the Gauss law 
constraint receives a contribution from the horizons of the background. For a similar 
result in the case of the free Maxwell field, we refer the reader to~\cite{Fernandes:2016imn} 
for spherically symmetric backgrounds and~\cite{Fernandes:2018pii} for a certain 
class of axisymmetric backgrounds. 
 
The action for scalar quantum electrodynamics on the 
spherically symmetric black hole spacetime is
\begin{equation}
S_{SQED} =  - \int  dV_4 \left(D_a \Phi (D_b \Phi)^* g^{ab} + m^2 \Phi\Phi^* + 
\tfrac{1}{4} F_{a b} F_{c d} g^{a c} g^{b d}\right) \, ,
\label{act.sqed}
\end{equation}
where $dV_4 = \lambda dV_x$ is the four dimensional volume form on the manifold
$\Sigma\times\mathbb{R}$ (with metric $g_{ab}$), $\Phi$ is a complex scalar field, 
$D_a = \partial_a + i g A_a$ is the gauge covariant derivative and 
$F_{a b} = 2 \partial_{[a} A_{b]}$ is the electromagnetic field strength tensor. 
We can now project this action on to the hypersurface described in the previous section. 
Time derivatives are given by the Lie derivative with respect to $\xi^a$, as in 
Eq.~(\ref{gen.dot}). In particular
\begin{equation}
\pounds_{\xi} a_b =  \dot a_b = -\lambda e_b + \mathcal{D}_b \phi \,,
\label{H.elf}
\end{equation}
where ${\mathcal{D}}_a$ is the spatial derivative defined in Eq.~(\ref{gen.den}) and
we have defined $e_d = - \lambda^{-1} \xi^c F_{c d}$\,.
By further defining $a_a = h_a^b A_b$,
$\phi = A_a \xi^{a}$, $\bar{D}_a = \partial_a + ig a_a$, $D_0 = \pounds_{\xi} + i g \phi$ 
and $f_{a b} = F_{c d} h^c_a h^d_b\,,$ we 
can rewrite Eq.~(\ref{act.sqed}) as 
\begin{equation}
S_{SQED} = \int dt \int \limits_{\Sigma} dV_x \,\lambda \left( \lambda^{-2} D_0 \Phi (D_0 \Phi)^* 
- h^{ab}\bar{D}_a\Phi \left(\bar{D}_b \Phi^*\right) - m^2 \Phi\Phi^* - \frac{1}{4} f_{a b} f^{ a b} 
+ \frac{1}{2} e_{a} e^{a} \right) \,.
\label{H.Lag}
\end{equation}
Denoting the conjugate momenta of $a_b\, , \phi \, , \Phi\,$ and $\Phi^*$ by 
$\pi^b, \pi^{\phi}\, , \Pi$ and $\Pi^*$ respectively, we have
\begin{align}
\pi^b &= \frac{\partial L_{SQED}}{\partial \dot{a}_b} = - e^{ b} \,, &  \pi^{\phi} &= 
\frac{\partial L_{SQED}}{\partial \dot{\phi}} = 0 \,, \notag\\
\Pi &= \frac{\partial L_{SQED}}{\partial \dot{\Phi}} =  \lambda^{-1} (D_0 \Phi)^* \,,  &
\Pi^* &= \frac{\partial L_{SQED}}{\partial \dot{\Phi}^*} =  \lambda^{-1} D_0 \Phi \,.
\label{H.mom}
\end{align} 
Thus the only primary constraint of the theory is 
\begin{equation}
\Omega_1 = \pi^{\phi}\,.
\label{U.con1}
\end{equation}
The canonical Hamiltonian can be constructed from the Legendre transform 
\begin{align} 
H_C &= \int \limits_{\Sigma} dV_x \, \left(\pi^b \dot{a}_b + \Pi \dot{\Phi} + \Pi^* \dot{\Phi}^* \right) -  L \notag \\
&= H_0 + \int \limits_{\Sigma} dV_x \,  \left(\pi^b \mathcal{D}_b \phi + i g \phi \left(\Phi^* \Pi^* - \Phi \Pi\right) \right)\,,
\end{align}
where $H_0$ is defined as
\begin{align}
H_0= \int \limits_{\Sigma} dV_x \, \lambda \left(\frac{1}{2} \pi^b \pi_b 
+ \frac{1}{4} f_{a b} f^{a b} + \Pi \,\Pi^* + m^2 \Phi \Phi^* + \bar{D}_a\Phi \left(\bar{D}^a\Phi\right)^* \right) \,.
\label{H0.brst}
\end{align}
Apart from the involvement of a non-flat manifold and its covariant derivatives, the 
definition of $H_0$ is the usual one known in flat spacetime which is used in the 
BRST treatment of this theory.

Using a multiplier 
$v_{\phi}$, we will now include the primary constraint $\pi^{\phi} \approx 0$ to the canonical Hamiltonian to define a new Hamiltonian
\begin{equation}
\widetilde{H}  = H_C + \int \limits_{\Sigma} dV_x \,  v_{\phi} \pi^{\phi}  \, .
\label{H.pri}
\end{equation}
The canonical Poisson brackets of the theory are
\begin{align}
\left[\phi(x), \pi^{\phi}(y) \right]_P &= \delta(x,y) \, , \quad & \quad 
\left[ a_b(x) , \pi^a(y) \right]_P  &= \delta^{a}_{b}\delta(x,y) \, , \\
\left[\Phi(x), \Pi(y) \right]_P &= \delta(x,y) \, , \quad & \quad \left[\Phi^*(x), \Pi^*(y)\right]_P  &= \delta(x,y) \, .
\label{H.can}
\end{align}
We now arrive at a key result used in this paper, 
namely that the Gauss law constraint of this theory is modified through the presence of horizons. 
This is determined by requiring that the primary constraint $\pi^{\phi}$ is satisfied at all times.
The consistency check of the primary constraint $\dot \pi^{\phi} \approx 0 \,$, is evaluated 
through the Poisson bracket of $\pi^\phi$ and $\tilde{H}$ with the help of a smearing function $\epsilon$ as follows,
\begin{align}
\int \limits_{\Sigma} dV_y \epsilon(y) \dot\pi^\phi(y) &= \int \limits_{\Sigma} dV_y 
\epsilon(y) \left[\pi^{\phi}(y) , \tilde{H} \right]_P  \,\notag \\
&= \int \limits_{\Sigma} dV_y 
\epsilon(y) \left[\pi^{\phi}(y) , \int \limits_{\Sigma} dV_x \pi^b(x) 
\mathcal{D}^x_b \phi(x) + i g \phi(x) \left(\Phi^*(x) \Pi^*(x) - \Phi(x) \Pi(x) \right) \right]_P \notag \\
&=  - \oint \limits_{\partial \Sigma} da_y \, \epsilon(y)  n^y_b \pi^b(y) + 
\int  \limits_{\Sigma} dV_y \, \epsilon(y) \left( \mathcal{D}^y_b \pi^b(y) - i g \left(\Phi^*(y) \Pi^*(y) - \Phi(y) \Pi(y) \right)\right)\,.
\label{U.PB1}
\end{align}
Here we have used the canonical Poisson brackets given in Eq.~(\ref{H.can}) 
and an integration by parts. The smearing function $\epsilon$ is assumed
to be well behaved, but $\epsilon$ or its first derivative are not required to vanish on the 
horizon (or horizons, if $\Sigma$ is the region between the horizons in a de Sitter 
black hole spacetime). By not requiring such conditions on the smearing functions at the horizons, 
we can equivalently state that we are not a priori adopting either Dirichlet or Neumann boundary conditions.
Then the Schwarz inequality demonstrates that the surface integral is finite
\begin{equation}
\left| n_b \pi^b \right|  \leq \, \sqrt{\left|n_b n^b\right| \, 
	\left|\pi_b \pi^b\right|}  \,.
\label{U.SI}
\end{equation}
In this expression, $n_bn^b = 1$ by definition, since $n_b$ is the `unit normal' to 
spatial sections of the horizon(s) and points in the direction of increasing time. 
Likewise, $\pi_b\pi^b = e_b e^b$ appears in the energy momentum tensor (more precisely 
in invariant scalars such as $T^{ab}T_{ab}$), and therefore may not diverge at the horizon.
It follows that the surface integral in Eq.~(\ref{U.PB1}) is finite at the horizon.
We can thus read off the Gauss law constraint from the last equality of Eq.~(\ref{U.PB1}), 
\begin{equation}
\Omega_2 =   n_b  \pi^b\Big\vert_{\cal{H}} - \mathcal{D}_b \pi^b + i g \left(\Phi^* \Pi^* - \Phi \Pi \right) \approx 0 \,.
\label{U.con2}
\end{equation}
The vertical bar $\Big\vert_{\cal{H}}$ on the first term in Eq.~(\ref{U.con2}) denotes that it is a contribution 
restricted to the horizon(s) of the spacetime. In other words, while the bulk contribution holds for all points 
of $\Sigma\,,$ the additional surface contribution of Eq.~(\ref{U.con2}) must be considered for all points at 
the horizons $\partial \Sigma$. Like the constraint $\pi^\phi\approx 0\,,$ this constraint also needs to be smeared 
with a well behaved function, regular at the horizons, for the purpose of Poisson bracket calculations.  We thus 
understand the constraint as 
\begin{equation}
\int \limits_{\Sigma} dV_x \epsilon(x) \Omega_2(x) =  \oint \limits_{\partial \Sigma} 
\epsilon(x) n^x_b  \pi^b(x) - \int \limits_{\Sigma} dV_x \,\epsilon(x) \left(\mathcal{D}^x_b \pi^b(x) 
- i g \left(\Phi(x)^* \Pi(x)^* - \Phi(x) \Pi(x) \right)\right) \approx 0 \,.
\label{U.con2int}
\end{equation}
Including the constraint Eq.~(\ref{U.con2}) in the Hamiltonian with its own multiplier $v_2$, we can write the total Hamiltonian as
\begin{equation}
H_T = H_0 + \int \limits_{\Sigma} dV_x \left((v_2 + \phi)\Omega_{2} + v_{\phi} \pi^{\phi}\right)\,.
\label{U.sham}
\end{equation}
It is straightforward to verify that $\dot{\Omega}_{2} \approx 0$, which reveals that there are no further constraints of the theory. 

Since $\left[\pi^{\phi}, \Omega_2\right]_P = 0$, the constraints are first class 
and generate gauge transformations of the fields. These transformations follow 
from the Poisson brackets of the fields with the general linear combination of 
the first class constraints, $\epsilon_1 \pi^{\phi} + \epsilon_2 \Omega_{2}$
\begin{align}
\delta \phi &= \epsilon_1  \, , \qquad \qquad \quad \, \delta a_b = \mathcal{D}_b \epsilon_2 \, , \notag\\
\delta \Phi &= - i g \epsilon_2 \Phi \, , \qquad \quad  \delta \Pi = i g \epsilon_2 \Pi \, , \notag\\
\delta \Phi^* &= i g \epsilon_2 \Phi^* \, , \qquad \quad \delta \Pi^* = - i g \epsilon_2 \Pi^*  \, .          
\label{SQED.gt}
\end{align}
We see that the gauge transformation of $a_b$, i.e. $\delta a_b$ in Eq.~(\ref{SQED.gt}), takes its usual form as on curved backgrounds without boundaries. This is a consequence of the surface term in the Gauss law constraint. We also note that the gauge transformations in Eq.~(\ref{SQED.gt}) hold for the fields throughout $\Sigma$, including the horizons. This in particular suggests the need to consider gauge fixing choices with surface terms at the horizons.

We can also determine the multipliers on the space of solutions of Hamilton's 
equations from the equations of motion. By considering $\left[\phi,H_T\right]_P$ 
we see that $v_{\phi} = \dot{\phi}$. Likewise, we note that $\left[a_b,H_T\right]_P$ gives the expression of 
Eq.~(\ref{H.elf}) provided $\partial_b v_2 =0$. This allows us to set $v_2 = 0$ without any loss of generality. 
With this choice for $v_{\phi}$ and $v_2$, we have
\begin{equation}
H_T = H_0 + \int \limits_{\Sigma} dV_x \left(\phi\Omega_{2} + \dot{\phi} \pi^{\phi}\right)\,.
\label{U.sham2}
\end{equation}
%
\subsection{Gauge fixing choices}

Before applying the Hamiltonian BRST formalism to this theory, 
it will be instructive to first describe how consistent gauge fixing choices can be determined within the Dirac-Bergmann 
formalism. Gauge fixing can be carried out through the introduction of additional 
constraints which have non-vanishing Poisson brackets with the first-class constraints of the theory. Given the two 
first-class constraints of the theory in Eq.~(\ref{U.con1}) and Eq.~(\ref{U.con2})
\begin{align}
\Omega_1 = \pi^{\phi}\,, \qquad \Omega_2 = n_b  \pi^b\Big\vert_{\cal{H}} - \mathcal{D}_b \pi^b + i g \left(\Phi^* \Pi^* - \Phi \Pi \right)\,,
\label{sqed.cons}
\end{align}
we need to introduce two additional constraints which should have non-vanishing Poisson brackets with those in 
Eq.~(\ref{sqed.cons}) and be consistent with them. The Gauss law constraint and its surface terms motivate the 
following gauge-fixing constraint
\begin{equation}
\Omega_3 = \mathcal{D}_b(\lambda^{-1} a^b) - n_b \lambda^{-1}a^b \Big\vert_{\cal{H}}\,.
\label{sqed.gf1}
\end{equation}
This constraint has the desired property of providing a non-vanishing Poisson bracket with $\Omega_2$. 
The consistency of this constraint requires that its time derivative with the Hamiltonian weakly vanishes. 
We find the following Poisson bracket of $\Omega_3$ with $H_T$ given in Eq.~(\ref{U.sham2})
\begin{equation}
\left[\Omega_3\,, H_T\right]_P = n_b  \pi^b\Big\vert_{\cal{H}} - \mathcal{D}_b \pi^b 
+ \left(n_b \lambda^{-1}\mathcal{D}^b\phi\right)\Big\vert_{\cal{H}} 
- \mathcal{D}_b \left(\lambda^{-1}\mathcal{D}^b\phi\right) \,.
\label{gf1.tevo}
\end{equation}
This weakly vanishes on account of $\Omega_2$, provided we impose
\begin{equation}
\Omega_4 = \mathcal{D}_b \left(\lambda^{-1}\mathcal{D}^b\phi\right) - \left(n_b \lambda^{-1}\mathcal{D}^b\phi\right)\Big\vert_{\cal{H}} + i g \left(\Phi^* \Pi^* - \Phi \Pi \right)\,.
\label{sqed.gf2}
\end{equation}
Thus $\Omega_3$ and $\Omega_4$ are consistent with the constraints of Eq.~(\ref{sqed.cons}) 
and have non-vanishing brackets with them. The construction of Dirac brackets on spherically 
symmetric backgrounds with horizons, following gauge fixing choices which involve surface terms 
at the horizons, has been described in further detail in~\cite{Fernandes:2016imn}. 
Eq.~(\ref{sqed.gf1}) and Eq.~(\ref{sqed.gf2}), apart from surface terms at the horizons, is the familiar choice of the radiation gauge for scalar electrodynamics~\cite{Sundermeyer:1982gv}. 

An alternative choice of gauge fixing, as far as surface terms at the horizons are concerned, is to consider 
null components of the fields at the horizons. This can be done by considering in place of 
Eq.~(\ref{sqed.gf1}) the following constraint
\begin{equation}
\Omega_3 = \mathcal{D}_b(\lambda^{-1} a^b) - \left(\lambda^{-1}n_b a^b - \lambda^{-2}\phi\right) \Big\vert_{\cal{H}}\,.
\label{sqed.gft1}
\end{equation}
Using Eq.~(\ref{nn.ss}), we recognize the surface term as a null component of $A_a$, specifically $\sqrt{2} \lambda^{-1} k_a A^a$.  We now find that $\dot{\Omega}_3$ weakly vanishes provided
\begin{equation}
\Omega_4 = \mathcal{D}_b \left(\lambda^{-1}\mathcal{D}^b\phi\right) - \left(\lambda^{-1} \left(n_b \mathcal{D}^b\phi - \lambda^{-1} v^{\phi}\right)\right)\Big\vert_{\cal{H}} + i g \left(\Phi^* \Pi^* - \Phi \Pi \right) \,.
\label{sqed.gft2}
\end{equation}
Since $v^{\phi} = \dot{\phi}$ on the space of solutions of Hamilton's equations, 
we can think of the surface term in Eq.~(\ref{sqed.gft2}) as equivalent to $\sqrt{2}\lambda^{-1} k^a \nabla_a \phi$. 
A few comments about the surface term in Eq.~(\ref{sqed.gft2}) may be in order. On the one hand, 
since the surface term contains the gauge dependent fields $a_b$ and $\phi$ and their derivatives, 
we cannot a priori impose any conditions on their finiteness be it at the horizon or elsewhere. 
On the other hand, we also note that regardless of the behaviour of $a_b$ and $\phi$ at $\cal{H}$, 
the null vectors $k^a$ and $l^a$ in Eq.~(\ref{nn.ss}) have a reparametrization invariance which can 
always be used to produce a finite result. For example, we can consider the change in normalization 
$k^a \to \lambda^{-1}k^a$ and $l^a \to \lambda l^a$ in Eq.~(\ref{nn.ss}), under which the relations 
in Eq.~(\ref{nn}) continue to hold. Thus the surface terms in Eq.~(\ref{sqed.gft1}) 
and Eq.~(\ref{sqed.gft2}) can always be adjusted so that they do not lead to divergences 
in any gauge-invariant quantity.

%

We thus see that fixing null components at the horizons is a consistent choice within the Dirac-Bergmann 
formalism and can admit an interesting set of Dirac brackets. In the following sections, 
we will derive results for the BRST invariant action where the effect of such gauge fixing 
choices will be shown to have more pronounced effects on the surface action at the horizon. 
We will also demonstrate that the use of Eq.~(\ref{sqed.gft1}) in the Hamiltonian BRST 
formalism can provide horizon corrections to the ghost number charge and 
dressed gauge invariant fields of scalar electrodynamics.
\section{Hamiltonian BRST formalism}\label{BRS}
We will now apply the Hamiltonian BRST formalism to derive the BRST invariant effective action and path integral 
for this theory. We will follow the standard treatment given in~\cite{Henneaux:1985kr, Henneaux:1992ig}, but on the 
black hole spacetimes with horizon contributions to the Gauss law and gauge fixing constraints, as described above.
We first extend the phase space of the previous section to include additional Grassmann odd fields, namely the 
ghosts and their momenta. Thus in addition to the fields considered in the previous section, we now introduce 
the ghost $\mathcal{C}$ and antighost $\bar{\mathcal{C}}$ and their conjugate momenta $\mathcal{P}\,, 
\bar{\mathcal{P}}\,,$ which satisfy
\begin{equation}
\left[\bar{\mathcal{P}}(x), \bar{\mathcal{C}}(y) \right]_P 
= - \delta(x,y) = \left[\mathcal{P}(x), \mathcal{C}(y) \right]_P \,.
\label{BRS.gh}
\end{equation} 
All brackets involving the ghosts other than those given in Eq.~(\ref{BRS.gh}) vanish. 
The ghost number can be determined from the ghost number charge
\begin{equation}
Q_C =  \int \limits_{\Sigma}dV_x \left(\mathcal{C} \mathcal{P} +  \bar{\mathcal{P}}\bar{\mathcal{C}} \right)\,.
\label{BRS.ghcharge}
\end{equation}
Given a functional of the fields $F$ in the extended phase space, we have
\begin{equation}
\left[F\,, Q_C\right]_P = \text{gh}(F) \, F \,,
\end{equation}
where $\text{gh}(F)$ denotes the ghost number of $F$. Then
\begin{align}
\text{gh}\left(\mathcal{C}\right) &= 1 = \text{gh}\left(\bar{\mathcal{P}}\right) \,, \notag\\
\text{gh}\left(\mathcal{P}\right) &= -1 = \text{gh}\left(\bar{\mathcal{C}}\right) \,.
\label{BRS.ghn}
\end{align}
Apart from the fields ${\mathcal{C}\,, \bar{\mathcal{C}}\,, \mathcal{P}\,, 
\bar{\mathcal{P}}\,,}$ all other canonical fields in the extended phase space have 
vanishing ghost number. 
%
%
%

The generator of BRST transformations $Q_{\text{BRST}}$ in the extended phase space can be directly constructed from the first-class constraints of a theory resulting from the Dirac-Bergmann formalism. Following the procedure in~\cite{Henneaux:1985kr} we have 
\begin{equation}
Q_{\text{BRST}} = \int \limits_{\Sigma} dV_x \left( \mathcal{C}(x) \Omega_{2}(x) - i \lambda \bar{\mathcal{P}}(x) \pi^{\phi}(x) \right) \, .
\label{BRS.gen}
\end{equation}
In Eq.~(\ref{BRS.gen}) we have included a factor of $\lambda$ in the second term. 
This is merely a convenient choice for what follows and does not result from more 
fundamental grounds. Nor does it contradict any known result, as $\lambda=1$ in flat spacetime.

The BRST charge is Grassmann odd and has ghost number $\text{gh}(Q_{\text{BRST}}) = 1$. 
BRST transformations of the fields are generated by its Poisson bracket with $Q_{\text{BRST}}$. Given a functional of the fields $F$, we will denote its BRST transformation by $s F$
\begin{equation}
s F = \left[F \, , Q_{\text{BRST}} \right]_P\, .
\label{brs.gi}
\end{equation} 
If $F$ has ghost number $\text{gh}(F)$ and mass dimension $d_F$, then $s F$ has ghost number $\text{gh}(F)+1$ and mass dimension $d_{F}+1$. By evaluating the Poisson brackets of the fields with $Q_{\text{BRST}}$ 
we find
\begin{align}
s a_b &= \mathcal{D}_b \mathcal{C} \,,  &  s \phi &= -i \lambda \bar{\mathcal{P}} \,,\notag\\
s \bar{\mathcal{C}} &= i \lambda \pi^{\phi} \,,  &  s \mathcal{P} &= -\Omega_2 \,,\notag\\
s \Phi &= - i g \mathcal{C} \Phi \,,  &  s \Pi &= i g \mathcal{C} \Pi \, , \notag\\
s \Phi^* &= i g \mathcal{C} \Phi^* \,,  &  s \Pi^* &= - i g \mathcal{C} \Pi^*  \,,\notag\\
s  \bar{\mathcal{P}} &= 0 = s \mathcal{C}\,, &
s \pi^{\phi} &=0= s \pi^a \,.
\label{BRS.tran}
\end{align}
Just as in the case of the gauge transformations generated by the first class constraints, the BRST transformations of the fields given in Eq.~(\ref{BRS.tran}) are the same as those on backgrounds without boundaries. 
The BRST charge is nilpotent, 
\begin{equation}
\left[Q_{\text{BRST}} \, ,Q_{\text{BRST}}\right]_P \equiv Q^2_{\text{BRST}} = 0 \,,
\label{brs.nil}
\end{equation} 
%
i.e., $s^2 F = 0$ for all $F$.

It is straightforward to verify that $H_0$ defined in Eq.~(\ref{H0.brst}) is invariant under the BRST transformations given in Eq.~(\ref{BRS.tran}). 
Further, since the BRST transformation is nilpotent, any BRST invariant quantity is known up to the addition 
of a term $ sF = \left[F \, , Q_{\text{BRST}} \right]_P$ for any $F$. 
In particular, we can define the following BRST invariant Hamiltonian  
\begin{equation}
H_{\text{BRST}} = H_0 - s\Psi \,, 
\label{bfv.ham}
\end{equation}
where $\Psi$ must have odd Grassmann parity and $\text{gh}(\Psi) = -1$, but 
can be arbitrary otherwise. As we will see shortly, $\Psi$ is used to implement 
gauge fixing choices within the Hamiltonian BRST formalism and hence is aptly 
known as the gauge fixing fermion.

%

From the Legendre transform with the Hamiltonian in Eq.~(\ref{bfv.ham}), we can also define the following BRST invariant action
\begin{equation}
S_{\text{BRST}} = \int dt \int \limits_{\Sigma} dV_x \left[\dot{a}_b \pi^b + \dot{\phi} \pi^{\phi} + \dot{\Phi}\Pi + \dot{\Phi}^* \Pi^* +  \dot{\bar{\mathcal{C}}} \bar{\mathcal{P}} + \dot{\mathcal{C}} \mathcal{P} - H_{\text{BRST}}\right] \, ,
\label{BRS.eff}
\end{equation}
%
Since $\Psi$ can be specified arbitrarily, physical processes are independent of the choice $\Psi\,,$ and hence independent of any gauge choice.
The invariance of the partition
function is expressed by the Fradkin-Vilkovisky theorem~\cite{Fradkin:1975cq,Batalin:1977pb}, which says that 
the path integral over all the canonical variables of the extended phase space $\mu^A \equiv \left(a_b\,, \pi^b\,, 
\phi \,, \pi^{\phi}\,, \Phi\,,\Pi\,,\Phi^*\,,\Pi^*\,,\mathcal{C}\,, \mathcal{P}\,, \bar{\mathcal{C}}\,,\bar{\mathcal{P}}\right)$
\begin{equation}
Z = \int \left[\mathcal{D} \mu^A \right] \text{exp}\left(i S_{\text{BRST}} \right) \,,
\label{brs.pathint}
\end{equation}
is independent of the choice of $\Psi$. 
The following $\Psi$ is customarily chosen
\begin{equation}
\Psi = \int \limits_{\Sigma} dV_x  \left(i\bar{\mathcal{C}}(x) \chi(x) + {\mathcal{P}}(x) \phi(x)\right)  \,,
\label{BRS.GFF}
\end{equation}
where $\chi$ is independent of the ghosts and their momenta, but can be specified arbitrarily otherwise. 
In the following we will also assume that $\chi$ is independent of all momenta other than $\pi^{\phi}$. 
Subject to this assumption, we now adopt Eq.~(\ref{sqed.gft1}) in the BRST 
formalism through the following choice of $\chi$
\begin{equation}
\chi = \mathcal{D}_a\left(\lambda^{-1} a^a\right) - \left(\lambda^{-1} n_a a^a - \lambda^{-2} \phi\right)\Big\vert_{\mathcal{H}} - \frac{1}{2}\pi^{\phi}\,,
\label{BRS.covchi}
\end{equation}
where as before, the symbol $\Big \vert_{\cal{H}}$ indicates that the 
term is evaluated at the horizons. Then the BRST transformation of $\Psi$ has the following expression
\begin{equation}
s \Psi = \int \limits_{\Sigma} dV_x \left(\lambda \pi^{\phi} \chi + i \lambda \mathcal{P} \bar{\mathcal{P}} + \phi \Omega_2 + i \bar{\mathcal{C}} s \chi\right) \,,
\label{BRS.psitran}
\end{equation}
where $\displaystyle{\int \limits_{\Sigma} dV_x \, i\bar{\mathcal{C}} s \chi}$ explicitly has the form
\begin{equation}
\int \limits_{\Sigma} dV_x \, i \bar{\mathcal{C}} s \chi = i \int \limits_{\Sigma} dV_x \, 
\bar{\mathcal{C}} \mathcal{D}_a \left(\lambda^{-1} \mathcal{D}^a \mathcal{C}\right) - 
i \oint \limits_{\partial \Sigma} da_x \, \lambda^{-1} \bar{\mathcal{C}} 
\left( n^a \mathcal{D}_a \mathcal{C} + i \bar{\mathcal{P}}\right) \,.
\end{equation}  
We can now use Eq.~(\ref{BRS.psitran}) to define $H_{\text{BRST}}$ and thus $S_{\text{BRST}}$, 
using which we have the following path integral from Eq.~(\ref{brs.pathint})  
\begin{align}
&Z = \int \left[\mathcal{D} \mu^A \right] \text{exp}\left(i S_{\text{BRST}} \right) \notag\\
& = \int \left[\mathcal{D} \mu^A \right] \text{exp} \Bigg(i \int dt \int \limits_{\Sigma} dV_x \Bigg[\dot{a}_a \pi^a + \dot{\phi} \pi^{\phi}  + \dot{\Phi}\Pi + \dot{\Phi}^* \Pi^* +  \dot{\bar{\mathcal{C}}} \bar{\mathcal{P}} + \dot{\mathcal{C}} \mathcal{P} - \phi \Omega_2 - i \bar{\mathcal{C}} s\chi \phantom{\int}  \notag\\
&  \qquad \quad \qquad \qquad - \lambda \left(\frac{1}{2} \pi^b \pi_b + \frac{1}{4} f_{a b} f^{a b} + \Pi \,\Pi^* + m^2 \Phi \Phi^* + \bar{D}_a\Phi \left(\bar{D}^a\Phi\right)^* + \pi^{\phi} \chi + i \mathcal{P} \bar{\mathcal{P}} \right) \Bigg] \Bigg)\,.
\label{BRS.pth}
\end{align}
We can now integrate out the momenta $\mathcal{P}$, $\bar{\mathcal{P}}$, $\Pi$, $\Pi^*$ and $\pi^a$ to find
\begin{align}
Z 
&= \int \left[\mathcal{D}a_a \,\mathcal{D}\phi \, \mathcal{D} \Phi \, \mathcal{D} \Phi^* \, \mathcal{D}\bar{\mathcal{C}}\, \mathcal{D}\mathcal{C}\, \mathcal{D} \pi^{\phi}\right] \text{exp} \left( i S_{\text{BRST}}\right)\,, 
\label{BRS.cov}\end{align}
where $S_{\text{BRST}}$ is now written as $S_{\text{BRST}} = S_{SQED} + S_{gh} + S_{gf}\,,$ with 
\begin{align}
&S_{SQED}= \int dt \int \limits_{\Sigma} dV_x \, \lambda \left( \frac{1}{2} e^a e_a - \frac{1}{4}f_{ab} f^{ab} + \lambda^{-2} D_0\Phi \left(D_0\Phi\right)^* - \bar{D}_a\Phi \left(\bar{D}^a\Phi\right)^* - m^2 \Phi \Phi^*\right)\,, \notag\\
&S_{gh} =  - i \int dt \int \limits_{\Sigma} dV_x \, \left(\lambda^{-1} \dot{\bar{\mathcal{C}}}\dot{\mathcal{C}} + \mathcal{D}_a \left(\lambda^{-1} \mathcal{D}^a \mathcal{C}\right) \right) - i \int dt \oint \limits_{\partial \Sigma} da_x\, \lambda^{-1} \bar{\mathcal{C}}\left(\lambda^{-1} \dot{\mathcal{C}} - n^{a} \mathcal{D}_{a} \mathcal{C}\right) \, , \notag\\
&S_{gf} = \int dt \int \limits_{\Sigma} dV_x \, \lambda \pi^{\phi} \left( \lambda^{-1}\dot{\phi} - \mathcal{D}_a \left(\lambda^{-1} a^a\right) + \frac{1}{2}\pi^{\phi}\right) - \int dt \oint \limits_{\partial \Sigma} da_x \,  \pi^{\phi} \left(\lambda^{-1}\phi - n^a a_a \right) \, .
\label{BRS.gfgh}
\end{align}
In deriving Eq.~(\ref{BRS.cov}), we performed a Gaussian integration over 
$\pi^a$ and made use of Eq.~(\ref{H.elf}). The integration over $\mathcal{P}\,,\bar{\mathcal{P}}\,, 
\Pi$ and $\Pi^*$ simply involve delta functions which enforce the relations $\bar{\mathcal{P}} = i \lambda^{-1} 
\dot{\mathcal{C}}$ and $\mathcal{P} = -i  \lambda^{-1} \left(\dot{\bar{\mathcal{C}}} +  
\left(\lambda^{-1} \bar{\mathcal{C}}\right)\Big\vert_{\cal{H}}\right)$, while $\Pi$ and 
$\Pi^*$ have their expressions given in Eq.~(\ref{H.mom}). 
Due to the surface term present in $S_{gh}$ in Eq.~(\ref{BRS.gfgh}), we cannot integrate out $\pi^{\phi}$ in the 
path integral as in the absence of a horizon, for example in flat spacetime. 
To identify the BRST transformations which leave $S_{SQED} + S_{gh} + S_{gf}$ in Eq.~(\ref{BRS.gfgh}) 
invariant, we can in Eq.~(\ref{BRS.cov}) simply substitute for all momenta other than $\pi^{\phi}$ 
their value at the extremum. This gives
\begin{align}
s a_b = \mathcal{D}_b \mathcal{C} \, , \quad & \qquad \quad s \phi = \dot{\mathcal{C}} \, ,\notag\\
s \Phi = - i g \mathcal{C} \Phi \, , \quad & \qquad \quad s \Phi^* = i g \mathcal{C} \Phi^* \, , \notag\\
s \bar{\mathcal{C}} &= i \lambda \pi^{\phi}  \,.
\label{BRS.tran2}
\end{align}
Likewise, we also find that the ghost number charge in Eq.~(\ref{BRS.ghcharge}) (or equivalently, the Noether charge from $S_{gh}$ in Eq.~(\ref{BRS.gfgh}) resulting from the scaling transformation $\bar{\mathcal{C}} \to e^{-s}\bar{\mathcal{C}}$ and $\mathcal{C} \to e^s \mathcal{C}$) now has the following expression
\begin{equation}
Q_C = i \int \limits_{\Sigma} dV_x \, \lambda^{-1} \left(\dot{\mathcal{C}}\bar{\mathcal{C}} - \mathcal{C}\dot{\bar{\mathcal{C}}}\right) - i \oint \limits_{\partial \Sigma} 
da_x \lambda^{-2} \mathcal{C}\bar{\mathcal{C}}\,.
\label{BRS.gc}
\end{equation}
We note that just as in the case of gauge dependent fields, we cannot assume any particular behaviour for the ghost fields on the background.
The surface integral in Eq.~(\ref{BRS.gc}) is absent in the ghost number charge on backgrounds without horizons. 
The implications of the surface integral for fields at the horizon can only be checked through the calculation of physical observables.
Our procedure seems to find explicitly the ghost degrees of freedom on the horizons of 
black holes. The presence of the horizon contributions to the ghost number charge in 
Eq.~(\ref{BRS.gc}) could be particularly relevant in the context of thermal gauge theories, 
as we will discuss later. 
%

\section{The co-BRST operator and dressed charges} \label{sec.cbrst}
We will now explore a construction in which the gauge fixing fermion $\Psi$ is the generator 
of nilpotent symmetry transformations and a conserved charge of the theory. We will follow the construction made 
previously in the context of quantum electrodynamics in flat spacetime~\cite{Lavelle:1993xf}. There
it was shown that a nilpotent operator $Q_{\text{BRST}}^{\perp}$\,, different from $Q_{\text{BRST}}$\,, exists 
which preserves the gauge fixing action and generates non-local and non-covariant transformations, that 
reduce the ghost number of the fields it acts on by one. It was also argued that 
physical states of the theory $\vert \Phi \rangle$ need to satisfy $Q_{\text{BRST}}\vert \Phi \rangle = 0$ 
and $Q_{\text{BRST}}^{\perp}\vert \Phi \rangle = 0$. 

Within the Hamiltonian BRST formalism, it was shown that this conserved 
charge can be identified with a gauge fixing fermion which generates the 
nilpotent transformations of $Q_{\text{BRST}}^{\perp}$ and which in addition can be be used to identify 
singlet states belonging to the BRST invariant inner product space~\cite{Marnelius:1997by}.
The gauge fixing fermion in this case is known 
as the co-BRST charge. In this subsection, we will demonstrate how we can choose a gauge fixing fermion with these 
properties, which will generalize the results of~\cite{Lavelle:1993xf,Marnelius:1997by} to curved backgrounds with horizons. 

We begin by noting that we are free to modify $H_0$ and $\Psi$ of Sec.~\ref{BRS} by the BRST differential
of an arbitrary functional $A$ as
\begin{align}
\widetilde{H}_0 &= H_0 + sA \notag\\
\widetilde{\Psi} &= \Psi + A \,,
\label{BRS.Hgftran}
\end{align} 
%
Under this modification, $H_{\text{BRST}}$ in Eq.~(\ref{bfv.ham}) remains invariant and the results 
in Sec.~\ref{BRS} are not affected. Let us thus consider the following expressions 
for $\widetilde{H}_0$ and $\widetilde{\Psi}$
\begin{align}
\widetilde{H}_0 &= H_0 - \int \limits_{\Sigma} dV_x \int \limits_{\Sigma} dV_y \frac{1}{2} \Omega_2(x) G(x,y) \Omega_2(y)  \notag\\
\widetilde{\Psi} &= \Psi + \int \limits_{\Sigma} dV_x \int \limits_{\Sigma} dV_y \frac{1}{2} \mathcal{P}(x) G(x,y) \Omega_2(y)   \,,
\label{BRS.Hgftran1}
\end{align} 
where $\Omega_2$ is as in Eq.~(\ref{U.con2}), $\Psi$ is as in Eq.~(\ref{BRS.GFF}) 
(with $\chi$ in $\Psi$ as in Eq.~(\ref{BRS.covchi})) and $G(x,y)$ is a Green function which satisfies
\begin{equation}
F(\mathcal{D})G(x,y) = - \delta(x,y)\,,
\label{fun.eq}
\end{equation}
with respect to a differential operator $F(\mathcal{D})$ which will be determined shortly. Since 
$s\mathcal{P} = -\Omega_2$ and $s\Omega_2 = 0\,,$ we see that $\widetilde{H}_0 - s \widetilde{\Psi} = H_0 - s \Psi$. 
%
%
Unlike $\Psi$, we can now show that $\widetilde{\Psi}$ can generate its own nilpotent symmetry transformations. We denote 
$\left[\mu^{\alpha}, \widetilde{\Psi}\right]_P = \bar{s} \mu^{\alpha}$ as the transformations generated by 
$\widetilde{\Psi}$, where $\mu^{\alpha} \equiv \left(a_a,\pi^a, \phi, \pi^{\phi}, \Phi, \Pi, \Phi^*, \Pi^*, \bar{\mathcal{C}},\bar{\mathcal{P}},\mathcal{C},\mathcal{P}\right)$ represents the set of all fields in the 
extended phase space. Evaluating the Poisson brackets of the fields with $\widetilde{\Psi}$ we find the 
following set of transformations
\begin{align}
\bar{s} a_b(x) &= \int \limits_{\Sigma}dV_y \frac{1}{2}\mathcal{P}(y) \mathcal{D}^x_b(G(x,y))\,,  &  \bar{s} \phi(x) &= - \frac{1}{2} i \bar{\mathcal{C}}(x)\,,\notag\\
\bar{s} \mathcal{C}(x) &= - \phi(x) - \int \limits_{\Sigma} dV_y \frac{1}{2}\Omega_2(y) G(x,y)\,,  & \qquad \qquad
 \bar{s} \bar{\mathcal{P}}(x) &= - i \chi(x)\,,\notag\\
\bar{s} \pi_a(x) &= i \lambda^{-1}(x) D_a^x \bar{\mathcal{C}}(x)\,,  &  \bar{s} \pi_{\phi}(x) 
&= - \mathcal{P}(x) - \left(i \lambda^{-2}(x)\bar{C}(x)\right)\Big\vert_{\mathcal{H}}\,, \notag\\
\bar{s} \Phi(x) &= - \int \limits_{\Sigma} dV_y \frac{1}{2} i g  \mathcal{P}(y) \Phi(x) G(x,y) \,,  &  
\bar{s} \Pi(x) &= \int \limits_{\Sigma} dV_y \frac{1}{2} i g \mathcal{P}(y) \Pi(x) G(x,y)\, , \notag\\
\bar{s} \Phi^*(x) &= \int \limits_{\Sigma} dV_y \frac{1}{2} i g \mathcal{P}(y) \Phi^*(x) G(x,y) \, ,  &  \bar{s} \Pi^*(x) &= - \int \limits_{\Sigma} dV_y \frac{1}{2} i g  \mathcal{P}(y) \Pi^*(x) G(x,y)  \,,\notag\\
& \hfill & \bar{s}  \bar{\mathcal{C}}(x) = 0 = \bar{s}\mathcal{P}(x) \, .
\label{cob.tran}
\end{align}
The Poisson bracket with $\widetilde{\Psi}$ reduces by 1 the ghost number of the field it acts on. The 
nilpotence of these transformations on all fields other than $\mathcal{C}$ and $\bar{\mathcal{P}}$ follow 
trivially. The transformations of $\mathcal{C}$ and $\bar{\mathcal{P}}$ are nilpotent provided the Green 
function $G(x,y)$ satisfies
\begin{equation}
\int \limits_{\Sigma} dV_x f(y) \mathcal{D}^x_a\left(\lambda^{-1}(x) \mathcal{D}_x^a G(x,y)\right) - \oint \limits_{\partial \Sigma} da_x \, f(y) n_x^a \lambda^{-1}(x) \mathcal{D}^x_a G(x,y) =  - f(x) \,,
\label{BFV.green}
\end{equation}
where $f(x)$ is any well behaved function on the hypersurface $\Sigma$. We can equivalently write Eq.~(\ref{BFV.green}) 
in the following way
\begin{equation}
\mathcal{D}^x_a\left(\lambda^{-1}(x) \mathcal{D}_x^a G(x,y)\right) - \left(\lambda^{-1} n_x^a \mathcal{D}^x_a G(x,y)\right)\Big \vert_{\cal{H}} = - \delta(x,y)\,.
\label{cob.gf}
\end{equation}
Thus we can write $F(\mathcal{D}) = \mathcal{D}_a\left(\lambda^{-1} \mathcal{D}^a\right) 
- \left(\lambda^{-1}n^a \mathcal{D}_a\right) \Big \vert_{\cal{H}}$. We note that the 
solution of $\mathcal{D}^x_a\left(\lambda^{-1}(x) \mathcal{D}_x^a G(x,y)\right) = - \delta(x,y)$ 
provides the electrostatic potential on spherically symmetric backgrounds~\cite{Cop:1928,J.Math.Phys.12.1845,Hanni:1973fn,Linet:1976sq,Leaute:1976sn}, 
whose algebraic expressions are known about the Schwarzschild~\cite{Linet:1976sq} and 
Reissner-Nordstr\"om spacetimes~\cite{Leaute:1976sn}.

Since the transformations
in Eq.~(\ref{BRS.Hgftran1}) do not affect the expression of $H_{\text{BRST}}$, the path integral 
in Eq.~(\ref{BRS.cov}) and the actions in Eq.~(\ref{BRS.gfgh}) are not modified.
One can now verify that $S_{\text{BRST}} = S_{SQED}+S_{gh}+S_{gf}$ is not only invariant under the BRST 
transformations given in Eq.~(\ref{BRS.tran2}), but also under the following co-BRST transformations
\begin{align}
\bar{s} a_b(x) &= - i \int \limits_{\Sigma}dV_y \frac{1}{2} \lambda^{-1}(y)\dot{\bar{\mathcal{C}}}(y) 
\mathcal{D}^x_b(G(x,y)) - i \oint \limits_{\partial \Sigma}da_y \frac{1}{2} \lambda^{-2}(y)\bar{\mathcal{C}}(y) 
\mathcal{D}^x_b(G(x,y)) \,, \notag\\
\bar{s} \Phi(x) &= - \int \limits_{\Sigma} dV_y \frac{1}{2} g  \lambda^{-1}(y)\dot{\bar{\mathcal{C}}}(y) \Phi(x) G(x,y) - \oint \limits_{\partial\Sigma} da_y \frac{1}{2}  g  \lambda^{-2}(y) \bar{\mathcal{C}}(y) \Phi(x) G(x,y) \,, \notag\\
\bar{s} \Phi^*(x) &=  \int \limits_{\Sigma} dV_y \frac{1}{2}  g \lambda^{-1}(y)\dot{\bar{\mathcal{C}}}(y) \Phi^*(x) G(x,y) + \oint \limits_{\partial\Sigma} da_y \frac{1}{2}  g  \lambda^{-2}(y) \bar{\mathcal{C}}(y) \Phi^*(x) G(x,y) \,, \notag\\
\bar{s} \mathcal{C}(x) &= \int \limits_{\Sigma} dV_y \frac{1}{2}\left(\mathcal{D}_a^y\left(\lambda^{-1}(y)\dot{a}^a(y)\right) -  i g  \lambda^{-1}(y) \left(\Phi^*(y) D^y_0\Phi(y) 
 - \Phi(y) D^y_0\Phi^*(y) \right)\right) G(x,y) \notag\\
& \qquad \qquad \qquad - \frac{1}{2}\phi(x) - \oint \limits_{\partial \Sigma} da_y \frac{1}{2} \lambda^{-1}(y) n_a^y \dot{a}^a(y) G(x,y) \, , \notag\\
\bar{s} \phi(x) &= - \frac{1}{2} i \bar{\mathcal{C}}(x) \, , \qquad \qquad \bar{s} \pi_{\phi}(x) = i\lambda^{-1}(x) \dot{\bar{\mathcal{C}}}(x) \,, \qquad \qquad \bar{s}  \bar{\mathcal{C}}(x) = 0 \,. 
\label{cob.tran2}
\end{align}
The transformations in Eq.~(\ref{cob.tran2}) were determined from Eq.~(\ref{cob.tran}) 
after substituting the values of the momenta at their extremum, specifically 
$\pi_a = \lambda^{-1}\dot{a}_a - \lambda^{-1}\mathcal{D}_a\phi$ and 
$\mathcal{P} = -i \lambda^{-1} \dot{\bar{\mathcal{C}}} - i \left(\lambda^{-2}
\bar{\mathcal{C}}\right)\Big\vert_{\cal{H}}$. These are the same expressions 
for the momenta as those in the previous section.
The co-BRST transformations in Eq.~(\ref{cob.tran2}) are the curved spacetime generalizations of those 
presented in~\cite{Lavelle:1993xf}, where they were used to demonstrate the invariance of dressed scalar fields. 
Specifically in flat spacetime, the dressed scalar fields defined as 
\begin{equation}
\Phi_{phys} = \Phi \, \text{exp} \left(-i g \frac{\partial_i A^i}{\nabla^2} \right)  \, , \qquad \Phi^*_{phys} = \Phi^* \, \text{exp} \left(  i g \frac{\partial_i A^i}{\nabla^2} \right) \,,
\label{sc.sc}
\end{equation}
satisfy the conditions $s \Phi_{phys} = 0 = \bar{s} \Phi_{phys}$ and $s \Phi^*_{phys} = 0 = \bar{s} \Phi^*_{phys}$\,,
while $\Phi$ and $\Phi^*$ do not. In Eq.~(\ref{sc.sc}), the index $i$ denotes spatial coordinates, 
while $\nabla^{-2}$ is the inverse Laplacian of flat spacetime which satisfies $\nabla_x^2 \nabla^{-2}
(\vec{x},\vec{y}) = -\delta(\vec{x}-\vec{y})$, where $\delta(\vec{x}-\vec{y})$ is the Dirac delta function 
on flat spacetime. Given the transformations in Eq.~(\ref{cob.tran2}) and the equation satisfied by the 
Green function $G(x,y)$ in Eq.~(\ref{cob.gf}), the following dressed fields
\begin{align}
\Phi_{phys}(x) &= \Phi(x) \, \text{exp} \left(- i g \int \limits_{\Sigma} dV_y \,\mathcal{D}_a^y\left(\lambda^{-1}(y) a^a(y)\right) G(x,y) + i g \oint \limits_{\partial \Sigma} da_y\, \lambda^{-1}(y) n_a^y a^a(y) G(x,y) \right)  \, , \notag\\
\Phi^*_{phys}(x) &= \Phi^*(x) \, \text{exp} \left(i g \int \limits_{\Sigma} dV_y \, \mathcal{D}_a^y\left(\lambda^{-1}(y) a^a(y)\right) G(x,y) - i g \oint \limits_{\partial \Sigma} da_y \, \lambda^{-1}(y) n_a^y a^a(y) G(x,y) \right)  \, .
\label{sc.sc2}
\end{align}
can be seen to satisfy $s \Phi_{phys} = 0 = \bar{s}\Phi_{phys}$ and $s \Phi^*_{phys} = 0 
= \bar{s}\Phi^*_{phys}$. In the flat limit, Eq.~(\ref{sc.sc2}) reduces to Eq.~(\ref{sc.sc}). 
The additional surface integrals now account for contributions from the horizons of the spacetime. 
In particular, the above expressions for dressed matter also hold for backgrounds with cosmological
horizons. Thus the co-BRST charge can be used to identify dressed matter fields on 
non-asymptotically flat black hole backgrounds. 

In the next section, we will discuss how the dressed fields in Eq.~(\ref{sc.sc2}) could be relevant in consideration of soft photon limits on backgrounds with horizons. Before proceeding to this discussion, we would like to provide a few comments on the above construction, specifically with regards to the choice of gauge. As we noted in Sec.~\ref{BRS}, the nilpotence of the BRST charge $Q_{\text{BRST}}$ allows us to choose $\Psi$ in any way we please. In the case of the co-BRST construction, an arbitrary choice can also be considered, but this would in general require modifying $\Omega_2$ in the expressions given in Eq.~(\ref{BRS.Hgftran1}) and the need for additional conditions on the Green function $G(x,y)$ apart from Eq.~(\ref{fun.eq}). In hindsight, we can state that the gauge as chosen in this section provides the simplest generalization of the known construction of the co-BRST charge in flat spacetime. Furthermore, as mentioned previously, in the absence of the surface terms in Eq.~(\ref{cob.gf}), the solution for the Green function is simply that of the electrostatic potential on spherically symmetric backgrounds, for which there exist known closed form expressions. Thus from the standpoint of determining physical observables from the path integral, the gauge considered in this paper will prove useful. As is well known, the results for physical observables will in any case be independent of the choice of gauge. 

\section{Summary and Discussion} \label{Con}
In this work, we considered the Hamiltonian BRST formalism for constrained theories on spherically symmetric 
backgrounds with horizons. We first provided the geometric framework needed to perform the Hamiltonian analysis 
on spacelike hypersurfaces orthogonal to the timelike Killing vector field of the spacetime. By considering 
the action for scalar quantum electrodynamics as an example, we then 
derived the constraints using the Dirac-Bergmann formalism. Keeping with our consideration of backgrounds with 
horizons, we were careful to evaluate Poisson brackets with smearing functions that are regular at the horizons. 
The Gauss law constraint, derived from the evaluation of Poisson brackets, was shown to involve terms from the 
horizon(s) of the spacetime. We then considered the Hamiltonian BRST formalism of the theory in the extended 
phase space involving the ghosts and their momenta. 
By fixing null components of the gauge fields at the horizons, we demonstrated that 
the ghost number charge of the theory involve additional corrections from the horizons 
of the background. We further considered gauge fixing fermions that 
generate their own nilpotent symmetry transformations which leave the action invariant. 
The gauge fixing fermion in this case is identified with the co-BRST operator. The 
requirement that physical fields are invariant under BRST and co-BRST transformations 
led us to identify dressed gauge invariant scalar fields of scalar electrodynamics, 
whose dressing function depends on the electromagnetic fields at the horizons of the background.

One of the avenues for further investigation following the results in this paper involves the 
quantization of gauge theories on black hole backgrounds. In identifying that the constraints are modified at 
the horizons, it is clear that as operator relations the constraints must be satisfied by states in the bulk and 
at the horizons. The mode expansion for gauge fields at the horizon and in particular their polarizations can be 
expected to be along the null directions at the horizons. This could be used to further explore the nature of 
``edge modes" at the horizon, along the lines of that which has been considered on the spatial boundaries of manifolds~\cite{Balachandran:1992qg,Balachandran:1995dv,Donnelly:2014fua,Blommaert:2018rsf,Blommaert:2018oue}. 
A consideration of the Hilbert spaces and the independent modes at the horizons and in the bulk of the spacetime 
lie outside the scope of the present work. We do note that in this regard the dressed gauge invariant fields, 
co-BRST operator and the corrections of the ghost number charge at the horizons, as considered in this paper, will be 
particularly useful. We should mention that a standard application of the BRST symmetry is its use in proving 
the renormalizability of a theory, using the Zinn-Justin equation for example. In our example above, the BRST transformations
on spacetimes with horizons turn out to be the same as on those without horizons, so the Zinn-Justin equation 
is unaffected.

The use of dressed gauge invariant fields in quantum electrodynamics was originally 
considered by Dirac~\cite{Dirac:1955uv}. The infrared properties of dressed fields 
were initiated in~\cite{Chung:1965zza,Kibble:1969} and their relevance in providing 
a finite S-matrix for quantum electrodynamics was provided by Faddeev and
 Kulish~\cite{Kulish:1970ut}. More recently, dressed fields have been shown to 
 provide a realization of the soft charges at null infinity on asymptotically 
 flat spacetimes which are consistent with Weinberg's soft photon theorem~\cite{Kapec:2017tkm}. 
 While soft hairs on the horizons of black holes of asymptotically flat spacetime have 
 been argued for in~\cite{Hawking:2016msc}, a similar realization of such soft hairs 
 in terms of dressed fields and their implications on black hole information remain 
 open problems. The dressed fields in Eq.~(\ref{sc.sc2}) could be useful in this respect. 

Ordinarily, there is a considerable amount of freedom in choosing the gauge dressing 
of fields in a given theory. For instance, the following dressed field in flat spacetime 
is perfectly legitimate
\begin{equation}
\Phi_{phys}(x) = \Phi(x) \text{exp}\left(\int \limits^x_{\Gamma} dz^i A_i(x_0,z)\right)\,,
\end{equation}
where the integral in the exponent is over some path $\Gamma$. This represents the 
`Wilson dressing' for a given scalar field $\Phi$. While such a dressing can be 
appropriate in the context of QCD and within holography, we note that in QED this 
dressing defines an infinitely excited state, where the electric flux is confined 
along $\Gamma$. On the other hand, the field given in Eq.~(\ref{sc.sc}) does provides 
the correct expression for the electric field of a static charge. It is particularly 
important to identify physically viable dressings in order to further investigate 
infrared properties and soft limits, which in the case of the dressed fields of 
Eq.~(\ref{sc.sc}) were studied in~\cite{Bagan:1999jf,Bagan:1999jk}. By involving 
horizon corrections to the dressing function of static scalar fields, 
Eq.~(\ref{sc.sc2}) in particular allows for the consideration of scattering processes 
near the horizon following the expansion of the exponential. 

The modification of the ghost number charge could also have interesting implications. 
This is particularly true for thermal gauge theories, whose partition function in the 
thermofield double formalism is known to depend on the ghost number 
charge~\cite{Hata:1980yr,Ojima:1981ma}. Specifically we note that while $\text{Tr} e^{-\beta H}$ 
provides the correct partition function for non-gauge theories, this is not 
the case in gauge theories whose state space involves unphysical degrees of 
freedom such as the longitudinal modes of the gauge fields and the ghosts. 
One can proceed to determine the physical state space either through the co-BRST 
construction or by adopting a special gauge, such as the Coulomb gauge and axial 
gauge, in which no physical particles appear. However a much simpler alternative 
was provided in~\cite{Hata:1980yr,Ojima:1981ma}, where it was shown that 
$\text{Tr}e^{-\beta H - \pi Q_C}$ describes the thermal partition function for 
gauge theories, consistent with the correct (periodic) boundary 
conditions of the ghosts. Thus by simply substituting $\text{Tr} e^{-\beta H}$ with 
$\text{Tr}e^{-\beta H - \pi Q_C}$, we can proceed with gauge theories just as one 
does in non-gauge theories. In the context of our paper, we demonstrated that both 
the Hamiltonian and ghost number charge involve surface corrections at the horizons 
of the background. This implies that known correlation functions and thermal 
propagators in flat spacetime could also be involve corrections from the horizons 
of the background. We look forward to performing these and related investigations in future work.


\end{document}